\documentclass[prd,aps,a4paper,superscriptaddress,twocolumn,nofootinbib]{revtex4}
\usepackage{graphicx}
\usepackage{color}
\usepackage{dcolumn}
\usepackage{bm}
\usepackage{slashed}
\usepackage{amsmath}
\usepackage{latexsym}
\usepackage{amssymb}
\usepackage{mathrsfs}
\usepackage{amsfonts}
\usepackage{url}
\usepackage{longtable}
\allowdisplaybreaks
\begin{document}
\title{The transverse-traceless gauge and the gauge problem of second order gravitational waves}

\author{Yadong Xue} 
\affiliation{Institute for Frontiers in Astronomy and Astrophysics, Beijing Normal University, Beijing 102206, China}
\affiliation{School of Physics and Astronomy, Beijing Normal University, Beijing 100875, China}
\author{Zhoujian Cao
\footnote{corresponding author}} \email[Zhoujian Cao: ]{zjcao@amt.ac.cn}
\affiliation{Institute for Frontiers in Astronomy and Astrophysics, Beijing Normal University, Beijing 102206, China}
\affiliation{School of Physics and Astronomy, Beijing Normal University, Beijing 100875, China}
\affiliation{School of Fundamental Physics and Mathematical Sciences, Hangzhou Institute for Advanced Study, UCAS, Hangzhou 310024, China}

\begin{abstract}
The gauge problem arises in the second order gravitational waves due to the mode mixing. Here, we introduce the transverse-traceless (TT) gauge to cosmological backgrounds, and find that if we choose the TT gauge at first order, the second order tensor mode would be gauge invariant. Analogous to the Ricci flat spacetime, the vacuum condition is the key to guarantee the existence of the TT gauge on cosmological backgrounds. When we have the vacuum condition, the Poisson gauge, the uniform curvature gauge, the synchronous gauge and the total matter gauge are all equivalent to the TT gauge. Once the vacuum condition is approximately satisfied, the Poisson gauge would reduce to the TT gauge at the same order of approximation. With the sub-horizon limit, the vacuum condition could be obtained approximately, and the Poisson gauge, the uniform curvature gauge and the synchronous gauge are all approximated TT gauge. Our findings explain several existing results in the literature and indicate that the proposed TT gauge is useful to discuss higher order gravitational waves.
\end{abstract}

\maketitle

\section{Introduction}

In general relativity, lower order perturbations will excite higher order perturbations according to Einstein equations. Among the metric perturbations, the tensor modes are referred as gravitational waves (GWs). Although the tensor modes are natively gauge-invariant at first order, they are not for second and higher orders \cite{matarrese1998relativistic,bruni1997perturbations}. As the most important example, the second order perturbations of the cosmological background do not admit a gauge invariant tensor modes. Consequently, a well-defined energy-momentum tensor for second order GWs still does not yet exist in cosmology. This issue is called the gauge problem of higher order GWs \cite{hwang2017gauge,domenech2021scalar,nakamura2019second,chang2020gauge,chang2020ongauge,chang2021note,de2020gauge,yuan2020scalar,kugarajh2025gauge,lu2020gauge,tomikawa2020gauge,inomata2020gauge,yuan2025gauge,ali2021gauge,domenech2021approximate,cai2022energy,ota2022covariant,comeau2023gauge,giovannini2020spurious,giovannini2020effective}, especially in cosmology.

The gauge problem of higher order GWs is firstly realized by Matarrese, Mollerach and Bruni in 1997 \cite{matarrese1998relativistic}. They blamed the issue to the mode mixing among the tensor modes, vector modes and scalar modes in a dust universe. In 2017, Hwang, Jeong and Noh \cite{hwang2017gauge} showed that the GW power spectrum strongly depends on the temporal gauge condition. For a more comprehensive review on the gauge problem history, one may refer to \cite{domenech2021scalar}.

After 2017, there have been many works dedicated to proposing possible solutions to the gauge problem \cite{nakamura2019second,chang2020gauge,chang2020ongauge,chang2021note,de2020gauge,yuan2020scalar,kugarajh2025gauge,lu2020gauge,tomikawa2020gauge,inomata2020gauge,yuan2025gauge,ali2021gauge,domenech2021approximate,cai2022energy,ota2022covariant,comeau2023gauge,giovannini2020spurious,giovannini2020effective}. Among them, some references \cite{nakamura2019second,chang2020gauge,chang2020ongauge,chang2021note} try to construct the gauge invariant second order tensor mode. However,
it was pointed out in \cite{chang2021note} that there are infinite families of gauge invariant constructions and one can not determine which one corresponds to the true GWs. Thus, it still requires further research.

A possible solution proposed in \cite{de2020gauge} is to find the most appropriate gauge describing the detection of GWs. On Ricci flat spacetimes, the transverse-traceless gauge is widely regarded as appropriate due to its computational convenience, as well as the unambiguous physical picture it provides for gravitational waves. The authors of Ref.~\cite{de2020gauge} argued that the synchronous gauge is the closest gauge to the TT gauge on cosmological backgrounds, therefore, the synchronous gauge seems to be a good gauge choice. Nonetheless, it was noted in \cite{lu2020gauge} that there exists residual gauge degrees of freedom in the synchronous gauge, which may lead to the gauge ambiguity. Different to Ref.~\cite{de2020gauge}, Ref.~\cite{giovannini2020effective} argued that the only reasonable gauges are those completely fixed.

Another commonly used gauge is the Poisson (Newtonian) gauge. After calculations, people found that the energy density spectra of second order GWs during radiation era is identical to the ones within Newtonian gauge, uniform curvature gauge and properly chosen synchronous gauge when the sub-horizon limit is satisfied \cite{yuan2020scalar,lu2020gauge,tomikawa2020gauge,inomata2020gauge,kugarajh2025gauge}. In the mean time, there also exists some gauges which result in different energy density spectra although on sub-horizon scales. Such gauges include the comoving orthogonal gauge, total matter gauge and others \cite{lu2020gauge,tomikawa2020gauge}. It seems that the sub-horizon limit could cause some gauge to have the same second-order GWs.

Following the viewpoint of \cite{matarrese1998relativistic}, the authors of Ref.~\cite{inomata2020gauge} argued that only the freely propagating tensor modes are true GWs. During radiation era, the second order tensor modes decouple from the scalar modes as they decays inside the horizon and finally freely propagate. Once the tensor modes are independent of the scalar modes, they will be gauge independent. In these three gauges, the freely propagating parts (true GWs) finally dominate the tensor modes, therefore, they have the same energy density spectra. Furthermore, the authors of Ref.~\cite{ali2021gauge} argued that if we only consider the tensor modes oscillating as $\sin(k\eta)/a$ or $\cos(k\eta)/a$ and drop all other terms, the left second order tensor modes are the same in seven various gauges whether they are during radiation or matter dominated era. Recently, Ref.~\cite{yuan2025gauge} tried to extend this viewpoint to any gauge through a boundary condition-based filtering method.

Combining the viewpoints of both \cite{de2020gauge} and \cite{inomata2020gauge}, Ref.~\cite{domenech2021approximate} showed that the second order GWs are approximately gauge invariant on sub-horizon scales as long as the source is not active and the spacetime slicing is well behaved. The authors suggested that the Newtonian gauge is a suitable gauge choice for the reason that it recovers the Newtonian gravity on small scales. Based on the Newtonian gauge, the second order GWs are approximately invariant under a set of reasonable gauge transformations.

There are also some references \cite{cai2022energy,ota2022covariant,comeau2023gauge,giovannini2020spurious,giovannini2020effective} trying to build a well-defined observable for second order GWs. In \cite{cai2022energy}, the energy of GWs was directly derived from the quasi-local gravitational energy \cite{wang2009quasilocal}. In Ref. \cite{ota2022covariant}, the authors claimed that the covariant TT part of the extrinsic curvature may represent the kinetic energy of the second order GWs. In Ref.~\cite{comeau2023gauge}, the author showed that the magnetic part of the Weyl tensor and the Cotton tensor of a slicing of spacetime are well-defined observables. Refs.~\cite{giovannini2020spurious,giovannini2020effective} argued that the pseudo energy momentum tensor of GWs with reasonable physical properties proposed in \cite{ford1977quantized} may solve the gauge ambiguity.

In the current paper, we follow the solution of Ref.~\cite{de2020gauge} to extend the TT gauge to the cosmological backgrounds. For Ricci flat spacetimes, the TT gauge is obtained based on the vacuum conditions \cite{flanagan2005basics,maggiore2008gravitational}. However, the cosmological background is instead always filled with matter. We alternatively introduce a `vacuum condition' to reduce the Poisson gauge, the uniform curvature gauge, the synchronous gauge and the total matter gauge to the TT gauge. And the sub-horizon limit could approximately reduce the Newtonian gauge, the uniform curvature gauge and the synchronous gauge to the TT gauge. In contrast, the sub-horizon limit can not reduce the total matter gauge to the TT gauge. This is because, on sub-horizon scales, the vacuum condition is only approximately valid, deviating from the idealized state of an exact vacuum. Meanwhile, we find that if we choose the TT gauge at first order, the second order tensor modes will be gauge independent. This answer the question why some gauges have the same energy density of GWs and while some not.

The current paper is organized as follows. In section \ref{TTgauge}, we propose to choose the TT gauge to treat the gauge problem. After that, we show that many commonly used gauges are equivalent to the TT gauge when the vacuum conditions is satisfied. In section \ref{approximateTT}, we discuss the situation that the vacuum condition is only approximately valid. In the limit where the vacuum condition becomes exact, the Newtonian gauge is found to approach the TT gauge correspondingly at the same order. And the sub-horizon limit means the vacuum condition is approximately satisfied. Our analysis makes several existing results in the literature can be easier understood. Finally, we conclude and discuss our main findings in the last section.

\section{The transverse-traceless gauge on the cosmological backgrounds}\label{TTgauge}

Consider a metric perturbation on the background spacetime, we can expand the component of metric tensor to second order
\begin{align}
    g_{\mu\nu}&=\bar{g}_{\mu\nu}+\lambda h_{\mu\nu}^{(1)}+\frac{1}{2}\lambda^2 h_{\mu\nu}^{(2)},
\end{align}
where a bar represents the background, the $h_{\mu\nu}^{(n)}(n=1,2)$ is $n$-th order perturbation and $\lambda$ is a dimensionless small parameter. For perturbed quantity, the gauge transformation comes from a second order infinitesimal coordinate transformation between the old (untilded) and new (tilde) coordinate system
\begin{align}
    x^\mu\rightarrow \tilde{x}^\mu=x^\mu+\lambda \xi^{\mu}_1+\frac{1}{2}\lambda^2 \xi^{\mu}_{2}.
\end{align}
where $\xi^{\mu}_n(n=1,2)$ is the $n$-th order expansion of $x^\mu$. According to the tensor transformation law, we have \cite{bruni1997perturbations}
\begin{align}
    \begin{aligned}
        \bar{g}_{\mu\nu}&\rightarrow \bar{g}_{\mu\nu},\\
        h_{\mu\nu}^{(1)}&\rightarrow h_{\mu\nu}^{(1)}+\mathscr{L}_{\xi^{(1)}}\bar{g}_{\mu\nu},\\
        h_{\mu\nu}^{(2)}&\rightarrow h_{\mu\nu}^{(2)}+\left(\mathscr{L}_{\xi^{(2)}}+\mathscr{L}_{\xi^{(1)}}^2\right)\bar{g}_{\mu\nu}+2\mathscr{L}_{\xi^{(1)}}h_{\mu\nu}^{(1)},
    \end{aligned}
    \label{htransformation}
\end{align}
where $\mathscr{L}$ is the Lie derivative operator. Here we have defined two infinitesimal coordinate transformation vectors
\begin{align}
    \xi^{(1)\mu}\equiv-\xi_1^\mu,\quad \xi^{(2)\mu}\equiv\xi^{\nu}_1\partial_\nu\xi^{\mu}_1-\xi_2^\mu.
\end{align}

In cosmology, a homogeneous, isotropic and spatially flat background is described by the Friedmann-Robertson-Walker metric
\begin{align}
    \mathrm{d}s^2=\bar{g}_{\mu\nu}\mathrm{d}x^{\mu}\mathrm{d}x^{\nu}=a^2\left[-\mathrm{d}\eta^2+\delta_{ij}\mathrm{d}x^i\mathrm{d}x^j\right],
\end{align}
where $\eta$ is the conformal time and $a=a(\eta)$ is the scale factor. In the current paper, a prime denotes differentiation with respect to the conformal time $\eta$ and $\mathcal{H}=a'/a$ is the comoving Hubble rate. The spacetime indices (Greek indices like $\mu,\nu,\cdots$) are raised and lowered with the background metric tensor $\bar{g}_{\mu\nu}$, and the spatial indices (Lattin indices like $i,j,k,\cdots$) are raised and lowered with three-dimensional Euclidean metric $\delta_{ij}$.

The $n$-th metric perturbation can be decomposed into scalar, vector and tensor modes \cite{stewart1990perturbations}
\begin{align}
    &h_{00}^{(n)}=a^2\left(-2\frac{1}{n!}\phi^{(n)}\right),\label{eq3}\\
    &h_{0i}^{(n)}=a^2\left[\frac{1}{n!}\left(B_{,i}^{(n)}-S_i^{(n)}\right)\right],\\
    &h_{ij}^{(n)}=a^2\left[\frac{1}{n!}\left(-2\psi^{(n)}\delta_{ij}+2E_{,ij}^{(n)}+2F_{(i,j)}^{(n)}+h_{ij}^{{(n)}\mathrm{TT}}\right)\right],\label{eq4}
\end{align}
where $\phi^{(n)}$, $\psi^{(n)}$, $B^{(n)}$ and $E^{(n)}$ are four scalar modes, $S_i^{(n)}$ and $F_{i}^{(n)}$ are two transverse (divergence-free) vector modes, and $h_{ij}^{{(n)}\mathrm{TT}}$ is the transverse-traceless tensor mode. The transverse-traceless tensor mode can be obtained directly through the transverse-traceless projector
\begin{align}
    &h_{ij}^{{(n)}\mathrm{TT}}=\mathcal{T}_{ij}^{lm} h_{lm}^{(n)}/a^2,\\
    &\mathcal{T}_{ij}^{lm}=(P^l{}_iP^m{}_j-\frac{1}{2}P_{ij}P^{lm}),\quad P^i{}_j\equiv\delta^i{}_j-\partial^i\Delta^{-1}\partial_j.\nonumber
\end{align}
Acting the transverse-traceless projector on Eq.~(\ref{htransformation}), we have the transformation rule for the $n$-th order transverse-traceless tensor mode
\begin{align}
    h_{ij}^{{(1)}\mathrm{TT}}&\rightarrow h_{ij}^{{(1)}\mathrm{TT}},\label{eq1}\\
    h_{ij}^{(2)\mathrm{TT}}&\rightarrow h_{ij}^{(2)\mathrm{TT}}+\mathcal{T}_{ij}^{lm}\left[\left(\mathscr{L}_{\xi_{(1)}}^2\bar{g}_{\mu\nu}+2\mathscr{L}_{\xi_{(1)}}h_{\mu\nu}^{(1)}\right)/a^2\right].\label{hijtt2tran}
\end{align}
Eq.~(\ref{eq1}) means that $h_{ij}^{{(1)}\mathrm{TT}}$ is gauge invariant. Eq.~(\ref{hijtt2tran}) indicates that $h_{ij}^{{(2)}\mathrm{TT}}$ is gauge dependent on and only on the first order gauge $\xi^{(1)\mu}$. If we fix the first order gauge while let higher order gauges free, $h_{ij}^{{(2)}\mathrm{TT}}$ does not change. We can call this property as restricted gauge invariant. In some sense, the restricted gauge invariant $h_{ij}^{{(2)}\mathrm{TT}}$ corresponds to the gauge invariant constructions for the second order tensor mode \cite{nakamura2019second,chang2020gauge,chang2020ongauge,chang2021note}. Similarly, this restricted gauge invariant $h_{ij}^{{(2)}\mathrm{TT}}$ does dependent on the first order gauge $\xi^{(1)\mu}$, so there are infinite families of choices. By analogy with Ricci flat spacetimes, we recommend the transverse-traceless (TT) gauge, and its convenience and physical significance will be  demonstrated. 

The first order TT gauge is defined as
\begin{align}
   h_{0\mu}^{(1)}=0,\quad h_{ij}^{(1)}=a^2 h_{ij}^{(1)\mathrm{TT}}.\label{eq2}
\end{align}
With this gauge, we can fix the first order gauge degrees of freedom $\xi^{(1)\mu}$ and remove the ambiguity of the second order tensor mode. In fact this idea can be extended to higher orders. Since $n$-th order metric perturbation only depends on lower order gauges, we can use the above idea to construct TT gauge for $n$-th order perturbation. Then order by order we can construct the whole TT gauge for each order perturbation.

The above mentioned TT gauge is ideal, but one caution is that it does not always exist. Nextly, the first order metric perturbation will be used as an example. For the sake of simplicity, we would ignore the order (1) on the first order perturbation, e.g., $\phi=\phi^{(1)}$. Only when necessary, the order (1) or (2) would be retained for distinction. In first order cosmological perturbation theory, it's convenient to use the three gauge invariant Bardeen variables \cite{bardeen1980gauge}
\begin{align}
    \Phi&=\phi-\mathcal{H}\sigma-\sigma^{\prime},\label{eq5}\\
    \Psi&=\psi+\mathcal{H}\sigma,\\
    \Xi_i&=F_i^{\prime}+S_i,\label{eq6}
\end{align}
where $\sigma=E^{\prime}-B$ is the shear potential. Within TT gauge, these Bardeen variables must vanish. On the other hand, these Bardeen variables are related to mater through Einstein equations.

Let us denote the mater energy momentum tensor as \cite{malik2009cosmological}
\begin{align}
    T_{\mu\nu}=pg_{\mu\nu}+(\rho+p)u_\mu u_\nu+\pi_{\mu\nu},
\end{align}
where $\rho$, $p$, $u_\mu$ and $\pi_{\mu\nu}$ are the energy density, isotropic pressure, the velocity four-vector and anisotropic stress tensor. Here, we only consider the first order matter perturbations
\begin{align}
    &\rho=\bar{\rho}+\delta \rho,\quad p=\bar{p}+\delta p,\nonumber\\
    &u_\mu=\bar{u}_\mu+\delta u_\mu,\quad \pi_{\mu\nu}=\bar{\pi}_{\mu\nu}+\delta \pi_{\mu\nu}.
\end{align}
where $\bar{u}_\mu=a(-1,0,0,0)$. For a homogeneous and isotropic background, we have $\bar{\pi}_{\mu\nu}=0$, therefore, the components of the energy momentum tensor on the background are
\begin{align}
    \bar{T}^0{}_0=-\bar{\rho},\quad \bar{T}^0{}_i=0,\quad \bar{T}^i{}_j=\delta^i{}_j\bar{p},
\end{align}
With the scalar, vector and tensor modes decomposition of the energy momentum perturbation
\begin{align}
    \begin{aligned}
        &\delta u^i=a^{-1}(\delta^{ij}v_{,j}+v^i),\\
        &\delta \pi_{ij}=a^2\left(\Pi_{,ij}-\frac13\nabla^2\Pi\delta_{ij}+\Pi_{(i,j)}+\Pi_{ij}\right),
    \end{aligned}   
\end{align}
the perturbed energy momentum tensor can be expressed as
\begin{align}
    \begin{aligned}
        &\delta T^0{}_0=-\delta\rho,\\
        &\delta T^0{}_i=(\bar{\rho}+\bar{p})(B_{,i}-S_i+v_{,i}+v_i),\\
        &\delta T^i{}_j=\delta p\delta^i{}_j+\delta \pi^i{}_j.
    \end{aligned}
\end{align}
Beside $\delta \pi_{ij}$, we could construct two more gauge invariant matter variables \cite{malik2009cosmological}
\begin{align}
    \bar{\rho}\Delta =\delta\rho+\bar{\rho}^{\prime}(v+B),\quad \delta q_i=(\bar{\rho}+\bar{p})(v_i-S_i),\label{eq14}
\end{align}
where $\Delta$ is called the comoving density contrast, and $\delta q_i$ is the transverse part of the 3-momentum.

With the above notations, the Einstein equations relating Bardeen variables and matter are
\begin{align}
    \nabla^2\Psi&=4\pi Ga^2\bar{\rho}\Delta,\label{eq19}\\
    \Psi-\Phi&=8\pi Ga^2\Pi,\\
    \nabla^2\Xi_i&=-16\pi Ga^2\delta q_i.
\end{align}
From these equations we can see if and only if
\begin{align}
    \bar{\rho}\Delta=\Pi=\delta q_i=0,\label{eq13}
\end{align}
the Bardeen variables vanish. That is to say $\bar{\rho}\Delta=\Pi=\delta q_i=0$ is the necessary condition for the existence of the TT gauge. We call this condition vacuum condition in the current paper. It should be noted that the vacuum conditions can be expressed using alternative sets of gauge invariant matter variables besides Eq.~(\ref{eq13}); however, these various expressions possess the same physical content.

In the next section, we will show that when the vacuum condition is satisfied, the Poisson gauge, the uniform curvature gauge, the synchronous gauge and the total matter gauge are all equivalent to the TT gauge. That means that the vacuum condition is both the sufficient and the necessary condition to guarantee the existence of the TT gauge.

\section{The relation between the TT gauge and the well known gauges} \label{gagues}

According to (\ref{eq3})-(\ref{eq4}), the TT gauge condition can be equivalently expressed with scalar, vector and tensor modes as
\begin{align}
    \phi=B=\psi=E=S_i=F_i=0.\label{eq7}
\end{align}

The Poisson gauge is defined as
\begin{align}
    B=0,\quad E=0,\quad F_i=0,
\end{align}
which generalizes the Newtonian gauge $B=E=0$ to include vector modes. Together with the definition of Bardeen variables (\ref{eq5})-(\ref{eq6}), we have
\begin{align}
    \begin{aligned}
        \Phi&=\phi,\\
        \Psi&=\psi,\\
        \Xi_i&=S_i.
    \end{aligned}  
\end{align}
So if only the vacuum condition is satisfied, the Bardeen variables vanish and the TT gauge (\ref{eq7}) is recovered.

The spatially flat or uniform curvature gauge is defined as
\begin{align}
    \psi=0,\quad E=0,\quad F_i=0.
\end{align}
Together with the definition of Bardeen variables (\ref{eq5})-(\ref{eq6}), we have
\begin{align}
    \begin{aligned}
        \Phi&=\phi+\mathcal{H}B+B^{\prime},\\
        \Psi&=-\mathcal{H}B,\\
        \Xi_i&=S_i.
    \end{aligned}
\end{align}
Again if only the vacuum condition is satisfied, the Bardeen variables vanish and the TT gauge (\ref{eq7}) is recovered.

The synchronous gauge is defined as
\begin{align}
    h_{0\mu}=0,
\end{align}
which is equivalent to
\begin{align}
    \phi=0,\quad B=0,\quad S_i=0.\label{eq11}
\end{align}
Decompose the infinitesimal coordinate transformation vector $\xi^{\mu}$ as
\begin{align}
    \xi^{\mu}=\left(\alpha,\; \delta^{ij}\beta_{,j}+\gamma^i\right),\quad \partial_i\gamma^i=0,
\end{align}
we have the following transformation rules for the metric perturbations
\begin{align}
    \begin{aligned}
        \phi&\to \phi+\mathcal{H}\alpha+\alpha^{\prime},\\
        \psi&\to \psi-\mathcal{H}\alpha,\\
        B&\to B-\alpha+\beta^\prime,\\
        E&\to E+\beta,\\
        S_i&\to S_i-\gamma_{i}',\\
        F_i&\to F_i+\gamma_i.
    \end{aligned}
\end{align}
Within the synchronous gauge conditions (\ref{eq11}), we still have the residual gauge degrees of freedom as 
\begin{align}
    \begin{aligned}
        &\alpha=\frac{C_1}{a},\\
        &\beta=C_2+C_1\int\frac{1}{a} d\eta,\\
        &\gamma_i=C_3,
    \end{aligned}
    \label{syncfreedom}  
\end{align}
where $C_{1,2,3}$ are three arbitrary constants. Meanwhile, the vanishing Bardeen variables (\ref{eq5})-(\ref{eq6}) lead us to
\begin{align}
    \begin{aligned}
        0&=\mathcal{H}E^{\prime}+E^{\prime\prime},\\
        0&=\psi+\mathcal{H}E^{\prime},\\
        0&=F_i^{\prime}.
    \end{aligned}
\end{align}
which can be solved out as 
\begin{align}
    \begin{aligned}
        &\psi=C_4\frac{\mathcal{H}}{a},\\
        &E=C_5-C_4 \int \frac{1}{a} d\eta ,\\
        &F_i=C_6,
    \end{aligned}
    \label{phiEF}  
\end{align}
where $C_{4,5,6}$ are three integral constants.
Combining  (\ref{syncfreedom}) and (\ref{phiEF}), we can choose 
\begin{align}
    C_1=C_4,\quad C_2=-C_5,\quad  C_3=-C_6,
\end{align}
to set $\psi=0$, $E=0$ and $F_i=0$, respectively. At this stage the synchronous gauge is the TT gauge.

The total matter gauge is defined as
\begin{align}
    B=-v,\quad E=0,\quad F_i=0.
\end{align}
Combining the above equations, the vacuum condition (\ref{eq13}) and the definition of the gauge invariant matter variables (\ref{eq14}), we have
\begin{align}
    \delta \rho =0,\quad  S_{i}=0,\quad \phi+\psi+B^{\prime}&=0.\label{eq16}
\end{align}
Based on Einstein equations, we have
\begin{align}
    \begin{aligned}
        &3\mathcal{H}(\psi^{\prime}+\mathcal{H}\phi)-\nabla^2(\psi+\mathcal{H}\sigma)=-4\pi Ga^2\delta\rho=0,\\
        &\psi^{\prime}+\mathcal{H}\phi=-4\pi Ga^2(\bar{\rho}+\bar{p})(B+v)=0,
    \end{aligned} 
\end{align}
which result in
\begin{align}
    &\psi^{\prime}+\mathcal{H}\phi=0,\label{eq15}\\
    &\psi-\mathcal{H}B=0.\label{phiB}
\end{align}
Plugging (\ref{phiB}) into (\ref{eq16}) and (\ref{eq15}), we get
\begin{align}
    &\phi+\mathcal{H}B+B^{\prime}=0,\label{eq17}\\
    &\mathcal{H}^{\prime}B+\mathcal{H}B^{\prime}+\mathcal{H}\phi=0.\label{eq18}
\end{align}
Combining (\ref{eq17}) and (\ref{eq18}), we get
\begin{align}
(\mathcal{H}^{\prime}-\mathcal{H}^2)B=0.
\end{align}
Excluding the special case with $\mathcal{H}=-(C+\eta)^{-1}$, we have $B=0$. And therefore $\phi=\psi=0$. So we find that the total matter gauge is the TT gauge when the vacuum condition (\ref{eq13}) is satisfied.

\section{The TT gauge and the sub-horizon limit}\label{approximateTT}

\subsection{The sub-horizon limit and the asymptotic vacuum conditions}

In cosmology, the second order GWs are produced by the primordial perturbations when they re-enter the horizon \cite{domenech2021scalar}. At this moment, the vacuum conditions are invalid, then the TT gauge does not exist. But this is natural, just like we can not talk about waves in source region. For the detection of these GWs, we concern about tensor modes well inside the horizon, where the sub-horizon limit $k\gg \mathcal{H}$ ($k$ is the wave number) is valid. In this section we will show that the sub-horizon limit can lead the vacuum conditions valid up to some order corrections of $\mathcal{H}/k$.

Current observation indicates that the scalar mode dominates the resulted first order metric perturbation. People accordingly only consider the scalar-scalar coupling plays a dominant role in the production of second order GWs, which is called the scalar-induced gravitational waves (SIGWs). We consider the same case in this section. Accordingly, the matter can be described by a perfect fluid ($\pi_{\mu\nu}=0$) with constant equation of state $w$, speed of sound $c_s^2$ and ignorable vector mode $v_i=0$.

Observations have revealed that the primordial perturbations are predominantly adiabatic on large scales which means that \cite{akrami2020planck1,akrami2020planck2}
\begin{align}
    w=c_s^2=\frac{\bar{p}}{\bar{\rho}}=\frac{\delta p}{\delta \rho}.
\end{align}
On such background, from the Friedmann equations, we get
\begin{align}
    a(\eta)\propto \eta^{1+b},\quad \mathcal{H}=(1+b)\eta^{-1},\quad b\equiv\frac{1-3w}{1+3w}.
\end{align}
Within the Newtonian gauge and according to Einstein equations, we have the master equation for $\phi$,
\begin{align}
    \phi^{\prime\prime}+3\mathcal{H}(1+c_s^2)\phi^{\prime}-c_s^2\nabla^2\phi=0.    \label{masterequation}
\end{align}
The general solutions to the master equation can be found in Ref.~\cite{baumann2007gravitational}. The behavior in the sub-horizon limit is \cite{domenech2021approximate},
\begin{align}
    \phi\propto(c_sx)^{-2-b},\quad c_s\neq0,
\end{align}
where $x\equiv k\eta$. It can be seen that the scalar mode $\phi$ decays as $x^{-2-b}$ on the sub-horizon scales except the matter dominated era $(w=c_s^2=0)$. Since $\pi_{\mu\nu}=0$, the Bardeen variables have relation $\Phi=\Psi=\phi$. Then Eq.~(\ref{eq19}) indicates that $\bar{\rho}\Delta$ decays as $x^{-2-b}$. That is to say the sub-horizon limit leads to the vacuum conditions up to the corrections of $O\left((\mathcal{H}/k)^{2+b}\right)$.

\subsection{The asymptotic vacuum conditions and the asymptotic TT gauge}

In the last section, we have shown that the Poisson gauge, the uniform curvature gauge, the synchronous gauge and the total matter gauge reduce to TT gauge when the vacuum condition is satisfied. However, when the vacuum condition is only approximately valid
\begin{align}
    \begin{aligned}
        &\bar{\rho}\Delta=O(x^{-\alpha}),\\
        &\Pi=O(x^{-\beta}),\\
        &\delta q_i=O(x^{-\gamma}),
    \end{aligned}
    \label{asymvacuum}
\end{align}
the above mentioned gauges are not always approximated TT gauges.

More specifically, under the asymptotic vacuum condition (\ref{asymvacuum}), the Bardeen variables admit behaviors
\begin{align}
    \begin{aligned}
        \Phi&=O(x^{-\alpha}),\\
        \Psi&=O(x^{-\min(\alpha,\beta)}),\\
        \Xi_i&=O(x^{-\gamma}).
    \end{aligned}   
\end{align}
Then the Poisson gauge has property
\begin{align}
    \begin{aligned}
        &B=E=F_i=0,\\
        &\phi=\Phi=O(x^{-\alpha}),\\
        &\psi=\Psi=O(x^{-\min(\alpha,\beta)}),\\
        &S_i=\Xi_i=O(x^{-\gamma}).
    \end{aligned}
\end{align}
That is to say the Poisson gauge asymptotically goes to TT gauge with the same order as the vacuum condition is satisfied.

Differently, it is not explicit for the relation between the TT gauge and the uniform curvature gauge, the synchronous gauge or the total matter gauge when the vacuum condition (\ref{eq13}) is replaced with the asymptotic vacuum condition (\ref{asymvacuum}). As an example to see such relations, we consider the case for SIGWs in last subsection again.

For the uniform curvature gauge, the only two nonvanishing scalar modes decay as
\begin{align}
    &B=-\Phi/ \mathcal{H} = O(x^{-1-b}),\\
    &\phi=2\Phi+(\Phi/ \mathcal{H})' =O(x^{-1-b}),
\end{align}
which means that the uniform curvature gauge approximately reduces to the TT gauge with the order of $O\left((\mathcal{H}/k)^{1+b}\right)$. The reduction speed is one order slower than the speed of the asymptotic vacuum condition.

For the synchronous gauge, the only two nonvanishing scalar modes behave as
\begin{align}
    \psi&=\Phi-\frac{\mathcal{H}}{a}\left(\int (-a\Phi) d\eta+C_1\right),\\
    E&=\int \left(\frac{\Phi-\psi}{\mathcal{H}}\right) d\eta +C_2.
\end{align}
where $C_1$ and $C_2$ depend on $k$ but are independent of $\eta$. According to Ref.~\cite{domenech2021approximate}, the scalar $\psi$ decays as $x^{-2-b}$, and $E$ behaves as
\begin{align}
    E \approx x^{-b} \left[\Phi(k)-C_1(k)\right]+O(x^{-2-b}).
\end{align}
By choosing $C_1(k)$ to counter $\Phi(k)$ \cite{domenech2021approximate}, the synchronous gauge could approximately reduce to the TT gauge with speed of $O\left((\mathcal{H}/k)^{2+b}\right)$.

For the total matter gauge, we have
\begin{align}
    \begin{aligned}
        &B=-\frac{\Phi^{\prime}+\mathcal{H}\Phi}{\mathcal{H}^{\prime}-\mathcal{H}^2}=O(x^{-b}),\\
        &\phi=\Phi-\mathcal{H}B-B^{\prime}=O(x^{-b}),\\
        &\psi=\Phi+\mathcal{H}B=O(x^{-1-b}).
    \end{aligned}
\end{align}
In general, the total matter gauge reduces to the TT gauge with the speed of $O\left((\mathcal{H}/k)^{b}\right)$, which is two orders slower than the speed of the asymptotic vacuum condition. For a special case when $b=0$, the TT gauge can not be asymptotically recovered at all.

It is shown in Refs.~\cite{yuan2020scalar,lu2020gauge,tomikawa2020gauge,inomata2020gauge,kugarajh2025gauge} that with the sub-horizon limit $k \gg \mathcal{H}$, the energy density for SIGWs yield the same prediction in the Poisson gauge, the uniform curvature gauge and the synchronous gauge during radiation dominated era. This result could be explained as that these three gauges approximately reduce to the TT gauge in the sub-horizon limit.

It was illustrated in Refs.~\cite{lu2020gauge,tomikawa2020gauge} that the total matter gauge has a second order tensor mode different from the one in the Newtonian gauge with the sub-horizon limit in a special case $b=0$. Our above analysis shows that the reason is that the total matter cannot approximately reduce to the TT gauge in the sub-horizon limit.

\section{Conclusion and discussion}\label{conclusion}

Due to mode mixing, the tensor modes are no longer gauge invariant for the second order gravitational waves. Specifically, the gauge transformation of $h_{ij}^{(2)\mathrm{TT}}$ depends on and only on the gauge transformation vector $\xi^{(1)\mu}$. Accordingly, if we fix the gauge at first order, i.e., let $\xi^{(1)\mu}=0$, then $h_{ij}^{(2)\mathrm{TT}}$ is gauge invariant.

Since the TT gauge has been regarded as the most suitable gauge for describing GWs on Ricci flat spacetimes \cite{maggiore2008gravitational}, it seems to be a good choice to start. If we choose the TT gauge at first order, then $h_{ij}^{(2)\mathrm{TT}}$ would be gauge invariant. Ref. \cite{de2020gauge,inomata2020gauge} stated that once the second order tensor modes have been decoupled from the source terms, and freely propagated to the detector, they would be gauge invariant. The vacuum condition is valid in this situation and the first order metric perturbation is already in the TT gauge. Based on our idea, the second order tensor mode is naturally gauge invariant.

The only limitation of TT gauge is that the vacuum condition must be satisfied, otherwise it does not exist. When the vacuum condition is satisfied, we interestingly find that many well known gauges including the Poisson gauge, the uniform curvature gauge, the synchronous gauge and the total matter gauge are equivalent to the TT gauge. When the vacuum condition is violated, the authors of \cite{domenech2021approximate} placed the Newtonian gauge on a priority position. A solid theoretical basis for such assumption is given in the current paper. We have shown that if only the vacuum condition is approximately satisfied, the Newtonian gauge would recover the TT gauge at the same approximating order.

Regarding the scalar induced gravitational waves (SIGWs), Refs.~\cite{yuan2020scalar,lu2020gauge,tomikawa2020gauge,inomata2020gauge,kugarajh2025gauge} demonstrated that the Poisson gauge, the uniform curvature gauge and the synchronous gauge have the same prediction in the sub-horizon limit $k \gg \mathcal{H}$. The authors in \cite{domenech2021approximate} explained that the SIGWs are approximately gauge invariant under a set of reasonable gauge transformations. From our perspective, the vacuum conditions holds approximately in the sub-horizon limit and these three gauges would reduce to the TT gauge approximately. Our analysis presents a simple explanation of these existing results.

Furthermore, our idea could be extended to higher orders. Since $n$-th order metric perturbation depends on and only on lower order gauges, we can construct TT gauge from the first to the $(n-1)$-th order. Then order by order, the whole tensor mode would be totally gauge invariant. Further studies on this generalization would be done in the future.

\section*{Acknowledgments}
This work was supported in part by the National Key Research and Development Program of China Grant No. 2021YFC2203001 and in part by the NSFC (No.~11920101003, No.~12021003 and No.~12005016). Z. Cao was supported by ``the Fundamental Research Funds for the Central Universities" of Beijing Normal University.

\bibliographystyle{unsrt}
\bibliography{refs}

\begin{thebibliography}{10}

\bibitem{matarrese1998relativistic}
Sabino Matarrese, Silvia Mollerach, and Marco Bruni.
\newblock Relativistic second-order perturbations of the einstein--de sitter universe.
\newblock {\em Physical Review D}, 58(4):043504, 1998.

\bibitem{bruni1997perturbations}
Marco Bruni, Sabino Matarrese, Silvia Mollerach, and Sebastiano Sonego.
\newblock Perturbations of spacetime: gauge transformations and gauge invariance at second order and beyond.
\newblock {\em Classical and Quantum Gravity}, 14(9):2585, 1997.

\bibitem{hwang2017gauge}
Jai-Chan Hwang, Donghui Jeong, and Hyerim Noh.
\newblock Gauge dependence of gravitational waves generated from scalar perturbations.
\newblock {\em The Astrophysical Journal}, 842(1):46, 2017.

\bibitem{domenech2021scalar}
Guillem Dom{\`e}nech.
\newblock Scalar induced gravitational waves review.
\newblock {\em Universe}, 7(11):398, 2021.

\bibitem{nakamura2019second}
Kouji Nakamura.
\newblock Second-order gauge-invariant cosmological perturbation theory: Current status updated in 2019.
\newblock {\em arXiv preprint arXiv:1912.12805}, 2019.

\bibitem{chang2020gauge}
Zhe Chang, Sai Wang, and Qing-Hua Zhu.
\newblock Gauge invariant second order gravitational waves.
\newblock {\em arXiv preprint arXiv:2009.11994}, 2020.

\bibitem{chang2020ongauge}
Zhe Chang, Sai Wang, and Qing-Hua Zhu.
\newblock On the gauge invariance of scalar induced gravitational waves: Gauge fixings considered.
\newblock {\em arXiv preprint arXiv:2010.01487}, 2020.

\bibitem{chang2021note}
Zhe Chang, Sai Wang, and Qing-Hua Zhu.
\newblock Note on gauge invariance of second order cosmological perturbations.
\newblock {\em Chinese Physics C}, 45(9):095101, 2021.

\bibitem{de2020gauge}
V~De~Luca, G~Franciolini, A~Kehagias, and A~Riotto.
\newblock On the gauge invariance of cosmological gravitational waves.
\newblock {\em Journal of Cosmology and Astroparticle Physics}, 2020(03):014, 2020.

\bibitem{yuan2020scalar}
Chen Yuan, Zu-Cheng Chen, and Qing-Guo Huang.
\newblock Scalar induced gravitational waves in different gauges.
\newblock {\em Physical Review D}, 101(6):063018, 2020.

\bibitem{kugarajh2025gauge}
Anjali~Abirami Kugarajh.
\newblock Gauge-dependence of scalar induced gravitational waves.
\newblock {\em Classical and Quantum Gravity}, 2025.

\bibitem{lu2020gauge}
Yizhou Lu, Arshad Ali, Yungui Gong, Jiong Lin, and Fengge Zhang.
\newblock Gauge transformation of scalar induced gravitational waves.
\newblock {\em Physical Review D}, 102(8):083503, 2020.

\bibitem{tomikawa2020gauge}
Keitaro Tomikawa and Tsutomu Kobayashi.
\newblock Gauge dependence of gravitational waves generated at second order from scalar perturbations.
\newblock {\em Physical Review D}, 101(8):083529, 2020.

\bibitem{inomata2020gauge}
Keisuke Inomata and Takahiro Terada.
\newblock Gauge independence of induced gravitational waves.
\newblock {\em Physical Review D}, 101(2):023523, 2020.

\bibitem{yuan2025gauge}
Chen Yuan, Yizhou Lu, Zu-Cheng Chen, and Lang Liu.
\newblock On the gauge invariance of secondary gravitational waves.
\newblock {\em Journal of Cosmology and Astroparticle Physics}, 2025(07):016, 2025.

\bibitem{ali2021gauge}
Arshad Ali, Yungui Gong, and Yizhou Lu.
\newblock Gauge transformation of scalar induced tensor perturbation during matter domination.
\newblock {\em Physical Review D}, 103(4):043516, 2021.

\bibitem{domenech2021approximate}
Guillem Dom{\`e}nech and Misao Sasaki.
\newblock Approximate gauge independence of the induced gravitational wave spectrum.
\newblock {\em Physical Review D}, 103(6):063531, 2021.

\bibitem{cai2022energy}
Rong-Gen Cai, Xing-Yu Yang, and Long Zhao.
\newblock On the energy of gravitational waves.
\newblock {\em General Relativity and Gravitation}, 54(8):89, 2022.

\bibitem{ota2022covariant}
Atsuhisa Ota, Hayley~J Macpherson, and William~R Coulton.
\newblock Covariant transverse-traceless projection for secondary gravitational waves.
\newblock {\em Physical Review D}, 106(6):063521, 2022.

\bibitem{comeau2023gauge}
Vincent Comeau.
\newblock Gauge-invariant scalar-induced gravitational waves from physical observables.
\newblock {\em arXiv preprint arXiv:2309.14624}, 2023.

\bibitem{giovannini2020spurious}
Massimo Giovannini.
\newblock Spurious gauge-invariance of higher-order contributions to the spectral energy density of the relic gravitons.
\newblock {\em International Journal of Modern Physics A}, 35(27):2050165, 2020.

\bibitem{giovannini2020effective}
Massimo Giovannini.
\newblock Effective anisotropic stresses of the relic gravitons.
\newblock {\em International Journal of Modern Physics D}, 29(16):2050112, 2020.

\bibitem{wang2009quasilocal}
Mu-Tao Wang and Shing-Tung Yau.
\newblock Quasilocal mass in general relativity.
\newblock {\em Physical review letters}, 102(2):021101, 2009.

\bibitem{ford1977quantized}
LH~Ford and Leonard Parker.
\newblock Quantized gravitational wave perturbations in robertson-walker universes.
\newblock {\em Physical Review D}, 16(6):1601, 1977.

\bibitem{flanagan2005basics}
Eanna~E Flanagan and Scott~A Hughes.
\newblock The basics of gravitational wave theory.
\newblock {\em New Journal of Physics}, 7(1):204, 2005.

\bibitem{maggiore2008gravitational}
Michele Maggiore.
\newblock {\em Gravitational waves}, volume~2.
\newblock Oxford university press, 2008.

\bibitem{stewart1990perturbations}
John~M Stewart.
\newblock Perturbations of friedmann-robertson-walker cosmological models.
\newblock {\em Classical and Quantum Gravity}, 7(7):1169--1180, 1990.

\bibitem{bardeen1980gauge}
James~M Bardeen.
\newblock Gauge-invariant cosmological perturbations.
\newblock {\em Physical Review D}, 22(8):1882, 1980.

\bibitem{malik2009cosmological}
Karim~A Malik and David Wands.
\newblock Cosmological perturbations.
\newblock {\em Physics Reports}, 475(1-4):1--51, 2009.

\bibitem{akrami2020planck1}
Yashar Akrami, Frederico Arroja, M~Ashdown, J~Aumont, Carlo Baccigalupi, M~Ballardini, Anthony~J Banday, RB~Barreiro, Nicola Bartolo, S~Basak, et~al.
\newblock Planck 2018 results-x. constraints on inflation.
\newblock {\em Astronomy \& Astrophysics}, 641:A10, 2020.

\bibitem{akrami2020planck2}
Yashar Akrami, F~Arroja, Mark Ashdown, J~Aumont, C~Baccigalupi, M~Ballardini, AJ~Banday, RB~Barreiro, N~Bartolo, S~Basak, et~al.
\newblock Planck 2018 results-ix. constraints on primordial non-gaussianity.
\newblock {\em Astronomy \& Astrophysics}, 641:A9, 2020.

\bibitem{baumann2007gravitational}
Daniel Baumann, Paul Steinhardt, Keitaro Takahashi, and Kiyotomo Ichiki.
\newblock Gravitational wave spectrum induced by primordial scalar perturbations.
\newblock {\em Physical Review D—Particles, Fields, Gravitation, and Cosmology}, 76(8):084019, 2007.

\end{thebibliography}

\end{document}